\documentclass[a4paper,twoside]{article}

\usepackage{epsfig}
\usepackage{subcaption}
\usepackage{calc}
\usepackage{amssymb}
\usepackage{amstext}
\usepackage{amsmath}
\usepackage{amsthm}
\usepackage{multicol}
\usepackage{pslatex}
\usepackage[english]{babel}
\usepackage{csquotes}
\usepackage[
    style=numeric,
    backend=biber,
    sorting=none,
    sortcites=true,
    url=true,
    eprint=true,
    giveninits=false,
    uniquename=true,
    minnames=1,
    maxnames=3
]{biblatex}

\AtEveryBibitem{%
  \clearfield{langid}%
  \clearfield{langidopts}%
  \clearfield{urldate}%

  \iffieldundef{doi}{}{\clearfield{url}\clearfield{urldate}}%
  \iffieldundef{eprint}{}{\clearfield{url}\clearfield{urldate}}%

  \clearname{editor}%
  \clearfield{series}%
  \clearfield{isbn}%
  \clearfield{issn}%
  \clearfield{day}%
  \clearfield{month}%
  \clearfield{place}%
  \clearlist{location}%
  \clearfield{eventtitle}%

  \ifentrytype{inproceedings}{\clearlist{publisher}}{}%

  \ifboolexpr{
    not test {\ifentrytype{software}}
    and
    not test {\ifentrytype{phdthesis}}
  }{%
    \clearfield{url}%
    \clearfield{urldate}%
  }{}%
}

\DeclareSourcemap{
  \maps[datatype=bibtex]{
    \map{
      \step[fieldset=urldate, null]
      \step[fieldset=language, null]
      \step[fieldset=langid, null]
      \step[fieldset=langidopts, null]
    }
  }
}

\usepackage{algorithm2e}
\usepackage[bottom]{footmisc}
\usepackage{SCITEPRESS}
\usepackage{color}
\usepackage{bm} 
\usepackage{overpic} 
\usepackage{hyperref}
\usepackage{physics} 
\usepackage{enumitem} 
\usepackage{booktabs}
\usepackage{tabularx}

\usepackage{cuted} 
\usepackage{caption} 

\definecolor{simonblue}{rgb}{0.22, 0.45, 0.70}
\definecolor{vietgreen}{rgb}{0.20, 0.60, 0.40}
\definecolor{richardkorange}{rgb}{0.93, 0.49, 0.19}

\usepackage{acronym}

\newacro{QST}{quantum state tomography}
\newacro{pQST}{parametric quantum state tomography}
\newacro{RBM}{restricted Boltzmann machine}
\newacro{FiLM}{feature-wise linear modulation}
\newacro{TFIM}{transverse-field Ising model}
\newacro{NQS}{neural-network quantum states}
\newacro{CD-$k$}{contrastive divergence}
\newacro{MPS}{matrix product state}
\newacro{ML}{machine learning}
\newacro{MLP}{multi-layer perceptron}
\newacro{ED}{exact diagonalization}
\newacro{MC}{Monte Carlo}
\newacro{QMC}{quantum Monte Carlo}
\newacro{KL}{Kullback--Leibler}
\newacro{SEM}{Standard Error of the Mean}

\newif\ifcamera
\cameratrue      

\begin{document}

\title{Parametric Quantum State Tomography with HyperRBMs}

\ifcamera
\author{\authorname{
    Simon Tonner\sup{1}\thanks{These authors contributed equally.}, 
    Viet T. Tran\sup{1}\footnotemark[1]\thanks{Corresponding author: viet\_thuong.tran@jku.at} 
    and Richard Kueng\sup{1}}
\affiliation{\sup{1}Department for Quantum Information and Computation at Kepler (QUICK), Johannes Kepler University, Linz, Austria}
}
\else
\author{\authorname{Anonymous Author(s)}}
\fi

\keywords{Quantum State Tomography, Machine Learning, Restricted Boltzmann Machines, Hypernetworks, Parametric Learning, Ising Model.}

\abstract{
Quantum state tomography (QST) is essential for validating quantum devices but suffers from exponential scaling in system size. Neural-network quantum states, such as Restricted Boltzmann Machines (RBMs), can efficiently parameterize individual many-body quantum states and have been successfully used for QST. However, existing approaches are point-wise and require retraining at every parameter value in a phase diagram. 
We introduce a parametric QST framework based on a hypernetwork that conditions an RBM on Hamiltonian control parameters, enabling a single model to represent an entire family of quantum ground states. Applied to the transverse-field Ising model, our HyperRBM achieves high-fidelity reconstructions from local Pauli measurements on 1D and 2D lattices across both phases and through the critical region. Crucially, the model accurately reproduces the fidelity susceptibility and identifies the quantum phase transition without prior knowledge of the critical point. These results demonstrate that hypernetwork-modulated neural quantum states provide an efficient and scalable route to tomographic reconstruction across full phase diagrams.
}
\onecolumn \maketitle \normalsize \setcounter{footnote}{0} \vfill

\section{\MakeUppercase{Introduction}}
\label{sec:introduction}

\begin{figure*}[!t]
    \centering
     \begin{overpic}[width=\linewidth]{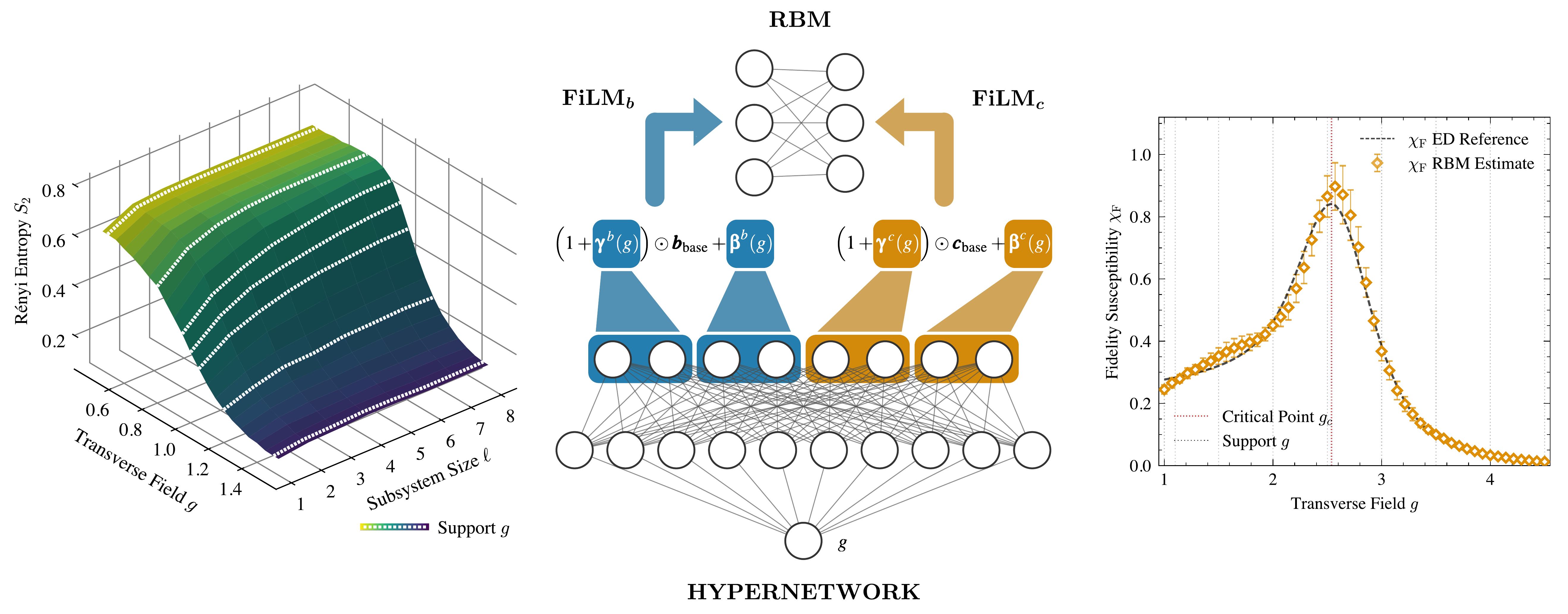}
            \put(3,5){\textbf{(a)}}
            \put(36,5){\textbf{(b)}}
            \put(73,5){\textbf{(c)}}
    \end{overpic}

    \caption{\textbf{HyperRBM reconstruction of Rényi entropy and susceptibility in the \acf{TFIM}.}
    (a) Second Rényi entropy $S_2(g)$ for the ground state of the $4\times4$ transverse-field Ising model for various subsystem sizes and transverse field strengths $g$. All results are obtained from a single HyperRBM trained on local Pauli measurements. Solid lines indicate exact values from \acl{ED}.  
    (b) HyperRBM architecture. A hypernetwork acts as conditioner and takes the control parameter $g$ and outputs \acl{FiLM} coefficients $\mathbf{\gamma}$ and $\mathbf{\beta}$, which affinely modulate the \acf{RBM} biases in visible and hidden layers.
    (c) Fidelity susceptibility $\chi_F(g)$ of the $4\times4$ TFIM.
Orange markers show HyperRBM estimates from free-energy gradient variance (Sec.~\ref{sec:fidelity_susceptibility}), gray dashed curves indicate ED references.
Vertical gray dotted lines mark support field strengths, and the red dotted line indicates the finite-size critical field $g_c$. The reconstructed $\chi_F(g)$ reproduces the ED peak in the crossover region.}
    \label{fig:main}
\end{figure*}

Extracting complete information about an unknown quantum state from measurement data requires the full reconstruction of a classical description of its wavefunction or density matrix. This process, known as \ac{QST} \cite{james_measurement_2001,paris_quantum_2004,banaszek_focus_2013}, is a fundamental tool across quantum science, enabling validation and benchmarking of quantum devices. The computational cost of \ac{QST} typically scales exponentially with system size, although polynomially scaling algorithms exist for certain restricted classes of states \cite{toth_permutationally_2010,schwemmer_experimental_2014}. However, the sample requirements of general \ac{QST} also scale exponentially. This holds even for the most sample-efficient tomographic protocols currently known \cite{DW16,HHJWN16,KRT17,FBK21,pelecanos_mixed_2025} and reflects a fundamental limitation of general-purpose \ac{QST}.

Shifting the goal from full state reconstruction to estimating specific properties allows for poly-logarithmic sample complexity. While highly efficient, this approach of \emph{shadow tomography} \cite{aaronson_shadow_2018,huang_predicting_2020,elben_randomized_2023} does not provide a full tomographic description of the quantum state.
Relatedly, a recent line of work studies how to predict ground-state properties of local, gapped Hamiltonians by learning from data drawn from other Hamiltonians in the same phase. 
These approaches leverage locality and classical-shadow style representations to obtain provable sample-complexity guarantees for estimating target observables within a phase, improving from polynomial to logarithmic and even constant sample complexity under additional assumptions \cite{huang_provably_2022,lewis_improved_2024,wanner_predicting_2024}. 
Despite these rigorous guarantees for learning specific observables, these methods do not aim to reconstruct a full state representation across the phase diagram. 

To achieve full state reconstruction with reduced complexity, alternative strategies incorporate physical assumptions. Methods such as \ac{MPS} tomography \cite{cramer_efficient_2010,baumgratz_scalable_2013} and compressed sensing \cite{gross_quantum_2010,KRT17} exploit low entanglement or low rank but are limited to specific classes of states. More recently, \ac{ML} techniques have emerged as powerful tools to address the curse of dimensionality in many-body quantum systems \cite{carleo_solving_2017}. \Acp{RBM} specifically have proven to be efficient ansatzes for \ac{QST} capable of reconstructing highly entangled states \cite{Torlai2017,nomura_restricted_2017}. In these \ac{NQS} representations, the wavefunction is compactly encoded in a network of stochastic neurons \cite{carleo_solving_2017}. Despite their success, existing neural \ac{QST} methods do not readily accommodate the study of phase diagrams, which are mapped out by varying a Hamiltonian control parameter. In practice, this requires training a separate model for each parameter value, resulting in significant computational and data inefficiencies. In this sense, current neural \ac{QST} approaches are inherently \emph{point-wise}, treating each parameter value as an independent learning problem and ignoring correlations between nearby states in parameter space.

Several approaches partially address this limitation. In one recent approach, \ac{pQST} is formulated as a classical post-processing problem, where the parameter dependence of the quantum state is assumed to admit an expansion in a fixed function basis \cite{schreiber_tomography_2025}. Standard point-wise tomography is performed at randomly sampled parameter values, and the resulting estimates are combined using compressed-sensing techniques to reconstruct the full parameterized family. In this approach, the parametric structure is fixed \emph{a priori}, and learning occurs only at the level of the expansion coefficients, rather than through a learned parametric state representation.

From a complementary perspective, conditional generative models have also been applied to quantum state tomography. Conditional generative adversarial networks \cite{ahmed_quantum_2021}, for example, have been used to learn a direct map from measurement data to a physical density matrix, enabling fast reconstruction across related experiments. While these methods introduce conditioning at the level of the generative model, they remain intrinsically point-wise: each reconstructed state is tied to a specific dataset and does not encode smooth parametric dependence on an underlying Hamiltonian. As a result, they do not provide access to quantities that rely on the continuity and differentiability of the state manifold, such as fidelity susceptibility.

A natural way to overcome these limitations is to learn the parametric dependence directly at the level of the quantum state ansatz itself. Hypernetworks \cite{ha_hypernetworks_2017,oswald_continual_2020,svensson_hyperpcm_2024}, which generate or modulate the parameters of another network as a function of external inputs, provide a principled mechanism for this purpose and have found success in meta-learning and conditional generative modeling \cite{munkhdalai_meta_2017,oswald_continual_2020,ratzlaff_hypergan_2019}. 
When applied to neural-network quantum states, this construction yields a differentiable family of wavefunctions conditioned on Hamiltonian parameters, enabling state reconstruction across an entire phase diagram. Crucially, treating the reconstructed states as a coherent parametric family allows us to extract the fidelity susceptibility directly from the learned model and to identify quantum phase transitions without prior knowledge of the critical point.

This motivates our work. We introduce a \ac{pQST} framework based on a hypernetwork-modulated \ac{NQS} that learns an entire family of ground states across a phase diagram within a single model. We build on the \ac{RBM}-based tomography of \cite{Torlai2017} by employing a lightweight hypernetwork implemented via \ac{FiLM} \cite{Perez2018} to condition the \ac{RBM} biases on an external Hamiltonian parameter.
We apply this Hyper\ac{RBM} architecture on the 1D and 2D \ac{TFIM} and demonstrate high fidelity reconstructions of ground states across both ferromagnetic and paramagnetic phases, including the critical region. Specifically, we consider a one-dimensional chain of size 16 and two-dimensional square lattices of size $L\times L$ with $L\in\{3,4\}$. Moreover, we show for the first time that an \ac{RBM}-based approach can predict the fidelity susceptibility, enabling the identification of the quantum phase transition without prior knowledge of the critical point \cite{zanardi_ground_2006,zanardi_information-theoretic_2007}. This demonstrates that \ac{pQST} captures not only the wavefunctions, but also higher-order parametric responses typically challenging for generative models.

While \ac{ML} approaches have been widely used for phase classification \cite{carrasquilla_machine_2017,van_nieuwenburg_learning_2017,kottmann_unsupervised_2020}, they focus on identifying phase boundaries rather than reconstructing quantum states. In contrast, our goal is a conditional generative model that captures the full parametric dependence of the state, enabling the evaluation of physical observables throughout the phase diagram. 

The paper is organized as follows. Section \ref{sec:theory} introduces the physical setting and reviews \acp{RBM} as neural-network quantum states. Section \ref{sec:methods} presents our Hyper\ac{RBM} architecture for parametric \ac{QST}. Section \ref{sec:evaluation} defines our evaluation metrics, covering both local observables and global state-level diagnostics. Section \ref{sec:experimental_setup} describes data generation and training details. Section \ref{sec:results} reports the main reconstruction results across the phase diagram. We conclude with a summary and outlook in Section \ref{sec:conclusion_outlook}.

\section{\MakeUppercase{Theory}}
\label{sec:theory}

\subsection{Neural-network quantum states and parametric tomography}

A central challenge in quantum many-body physics is the exponential growth of Hilbert space dimension. A generic wavefunction of $N$ spins,
\begin{equation}
|\Psi\rangle = \sum_{\boldsymbol{s}} \Psi(\boldsymbol{s})\,|\boldsymbol{s}\rangle ,
\end{equation}
requires $2^N$ complex-valued coefficients in a fixed product basis $\{|\boldsymbol{s}\rangle\}$ with $\boldsymbol{s} \in \left\{0,1\right\}^N$. Neural-network quantum states (NQS) address this issue by replacing the explicit storage of amplitudes with a parameterized function that approximates the mapping
$\boldsymbol{s} \mapsto \Psi(\boldsymbol{s}) \in \mathbb{C}$.

In quantum state tomography, the wavefunction is not given explicitly but accessed through quantum measurement data. We consider a \emph{parametric} tomography setting in which the quantum state depends on an external control parameter $g$. This parameter may represent a Hamiltonian control variable (e.g., transverse field strength). The goal is to learn a single conditional model $p_\theta(\cdot\mid g)$ that captures a continuous family of quantum states from measurement data collected at multiple parameter values.

For general quantum states, reconstructing $\Psi(\boldsymbol{s})$ requires learning both amplitudes and phases and typically necessitates measurements in multiple bases. In this work, we focus on ground states of \emph{stoquastic} Hamiltonians. As a consequence of the Perron\nobreakdash-Frobenius theorem, these states can be chosen to be real-valued and non-negative in the fixed product basis, i.e.\ $\Psi(\boldsymbol{s}\mid g)\ge 0$ for all $\boldsymbol{s}$ \cite{bravyi_complexity_2008}.
In this setting, the quantum state is fully characterized by the probability distribution of measurement outcomes in the computational ($\sigma^z$) basis,
\begin{equation}
q(\boldsymbol{s}\mid g) = |\Psi(\boldsymbol{s}\mid g)|^2 ,
\end{equation}
and quantum state reconstruction reduces to learning this distribution from data. The training dataset consists of tuples $(\boldsymbol{s}^{(i)}, g^{(i)})$ collected at multiple values of $g$. In the following we realize the conditional model $p_\theta(\boldsymbol{s}\mid g)$ with a \ac{RBM}.

\subsection{Restricted Boltzmann machines}
For completeness and to make the later conditioning mechanism explicit, we begin by briefly reviewing the standard \ac{RBM} formulation for modeling a single quantum state and highlight the parameters (biases) that will be modulated by the \ac{FiLM} hypernetwork in Sec.~\ref{sec:methods}, before extending it to a conditional, parameter-dependent model in Sec.~\ref{sec:conditional_energy}.

To represent probability distributions over high-dimensional binary variables, we employ a \ac{RBM} \cite{hinton_training_2002,hinton_practical_2012} which is a bipartite energy-based model defined on a set of binary visible units
$\boldsymbol{v} = (v_1,\dots,v_N)$ and binary hidden units
$\boldsymbol{h} = (h_1,\dots,h_M)$.
The joint distribution is given by the Boltzmann form
\begin{equation}
p_\theta(\boldsymbol{v},\boldsymbol{h})
= \frac{1}{Z_\theta}\exp\!\left[-E_\theta(\boldsymbol{v},\boldsymbol{h})\right],
\end{equation}
with energy function
\begin{equation}
E_\theta(\boldsymbol{v},\boldsymbol{h})
= -\sum_i b_i v_i - \sum_j c_j h_j - \sum_{i,j} v_i W_{ij} h_j , \label{eq:rbm_energy}
\end{equation}
where $\theta = \{\boldsymbol{W},\boldsymbol{b},\boldsymbol{c}\}$ denotes the set of trainable parameters and $Z_\theta=\sum_{\boldsymbol{v},\boldsymbol{h}} e^{-E_\theta(\boldsymbol{v},\boldsymbol{h})}$ is the partition function. 
The bias terms $\boldsymbol{b}$ and $\boldsymbol{c}$ enter additively in the energy and will be the quantities modified in our parametric extension (see Sec.~\ref{sec:hypernetwork}).

Marginalizing over the hidden variables $\boldsymbol{h}$ yields the model distribution over visible configurations
\begin{equation}
p_\theta(\boldsymbol{v})
= \frac{1}{Z_\theta}\sum_{\boldsymbol{h}} e^{-E_\theta(\boldsymbol{v},\boldsymbol{h})}
= \frac{1}{Z_\theta} e^{-F_\theta(\boldsymbol{v})}.\label{eq:boltzmann_prob_v}
\end{equation}

Here, $F_\theta (\boldsymbol{v})$ is the free energy given by
\begin{equation}
F_\theta(\boldsymbol{v})
= - \sum_i b_i v_i
  - \sum_j \mathrm{softplus}\!\left( (\boldsymbol{W}^{\mathsf T}\boldsymbol{v})_j + c_j \right), \label{eq:rbm_free_energy}
\end{equation}
where $\mathrm{softplus}(x)=\log\!\left(1+e^{x}\right)$ arises from analytically summing over all binary hidden units.

The bipartite structure implies conditional independence between visible and hidden variables,
allowing efficient block Gibbs sampling and tractable evaluation of
$F_\theta(\boldsymbol{v})$, which is the central quantity entering training,
sampling, and observable estimation.

\subsection{Training and contrastive divergence}

\ac{RBM} training is formulated as the minimization of the \ac{KL} divergence between the empirical data distribution $q(\boldsymbol{v})$ and the model distribution $p_\theta(\boldsymbol{v})$:
\begin{equation}
D_{\mathrm{KL}}\!\left(q \,\|\, p_\theta\right)
= \sum_{\boldsymbol{v}} q(\boldsymbol{v}) \log \frac{q(\boldsymbol{v})}{p_\theta(\boldsymbol{v})}.
\end{equation}
Taking the gradient with respect to $\theta$ and noting that the entropy of $q$ is independent of the model parameters, one finds
\begin{align}
\nabla_\theta D_{\mathrm{KL}}
&= - \sum_{\boldsymbol v} q(\boldsymbol v)\,\nabla_\theta \log p_\theta(\boldsymbol v) .
\end{align}
Using $\log p_\theta(\boldsymbol{v}) = -F_\theta(\boldsymbol{v}) - \log Z_\theta$, the gradient of the log partition function can be expressed as an expectation value:
\begin{align}
\nabla_\theta \log Z_\theta
&= \frac{1}{Z_\theta}\sum_{\boldsymbol v} e^{-F_\theta(\boldsymbol v)} \left( - \nabla_\theta F_\theta(\boldsymbol v) \right) \nonumber \\
&= -\mathbb{E}_{p_\theta}\!\left[\nabla_\theta F_\theta(\boldsymbol v)\right].
\end{align}
Substituting this back yields the gradient of the KL divergence as:
\begin{equation}
\label{eq:grad_kl}
\nabla_\theta D_{\mathrm{KL}}
= \mathbb{E}_{q}\!\left[\nabla_\theta F_\theta(\boldsymbol{v})\right]
- \mathbb{E}_{p_\theta}\!\left[\nabla_\theta F_\theta(\boldsymbol{v})\right].
\end{equation}
The first term, known as the positive phase, lowers the free energy of configurations observed in the data. The second term, the negative phase, raises the free energy of configurations sampled from the model.
While the positive phase is evaluated directly on data samples, the negative phase requires sampling from $p_\theta(\boldsymbol{v})$. Exact sampling is intractable for large systems; therefore, the expectation $\mathbb{E}_{p_\theta}$ is approximated using \ac{CD-$k$} \cite{hinton_training_2002}. In this scheme, short Gibbs chains are initialized at data configurations to obtain samples $\boldsymbol{v}_{CD}$.

Finally, the model parameters are updated using stochastic gradient descent. Based on the approximated gradient from Eq.~\eqref{eq:grad_kl}, the update rule is given by:
\begin{equation}
\theta^{(t+1)} \leftarrow \theta^{(t)} - \eta \Big( \mathbb{E}_{q}\!\left[\nabla_\theta F_\theta(\boldsymbol{v})\right] - \mathbb{E}_{p_{CD}}\!\left[\nabla_\theta F_\theta(\boldsymbol{v}_{CD})\right] \Big), \label{eq:grad_update}
\end{equation}
where $\eta$ is the learning rate.

\subsection{Application to Ising-type ground states}

We now connect the \ac{RBM} formalism to quantum spin systems. For the transverse-field Ising model, the visible units are identified with spin configurations in the computational basis. To align with the physical spin values, we formulate the \ac{RBM} visible units as $\boldsymbol{v} \equiv \boldsymbol{s} \in \{0,1\}^N$.

Since the ground states considered here are known to be real-valued and non-negative (due to the stoquasticity of the Hamiltonian), the trained \ac{RBM} parametrizes the wavefunction amplitude directly:
\begin{equation} 
    \Psi_\theta(\boldsymbol{s}) = \sqrt{p_\theta(\boldsymbol{s})}. \label{eq:pos_wavefunction_ansatz}
\end{equation} 

While we restrict our attention here to stoquastic Hamiltonians to isolate the problem of learning the parametric dependence on $g$, the hypernetwork formalism can be extended to complex-valued states by using two (Hyper)RBMs to parametrize the amplitude and phase separately, as demonstrated in Ref.~\cite{Torlai2017}.

\section{\MakeUppercase{Methods}}
\label{sec:methods}

\subsection{Problem setup: Transverse-field Ising model}
\label{sec:tfim_setup}

We study ground states of \ac{TFIM} in one and two spatial dimensions. The Hamiltonian is given by
\begin{equation}
H(g) = -J \sum_{\langle i,j\rangle} \sigma_i^z \sigma_j^z
       - g \sum_{i=1}^{N} \sigma_i^x,
\label{eq:tfim_hamiltonian}
\end{equation}
where, $J>0$ is the ferromagnetic coupling and $g$ is the transverse-field strength, serving as our continuous control parameter.

We consider two distinct geometries with periodic boundary conditions:
\begin{enumerate}[label=(\roman*)]
    \item \textbf{1D Chain:} A single one-dimensional chain consisting of $N=16$ spins. 
    \item \textbf{2D Lattices:} Two-dimensional square lattices of size $L\times L$ with $L\in\{3,4\}$ corresponding to $N=9$ and $N=16$ spins, respectively.
\end{enumerate}

For a set of support points $\{g_i\}$, we collect measurement samples $\{(\boldsymbol{s}^{(j)}, g^{(j)})\}$ in the computational basis. The goal is to train a single conditional model $p_\theta(\boldsymbol{s}\mid g)$ that accurately interpolates across the entire range of $g$ for each geometry. Reference ground states $|\Psi_{\mathrm{ED}}(g)\rangle$ are computed via Lanczos \ac{ED} for benchmarking.

\subsection{Conditional RBM energy}
\label{sec:conditional_energy}

We model the conditional distribution $p_\theta(\boldsymbol{s}\mid g)$ using
an \ac{RBM} as introduced in Sec.~\ref{sec:theory}.
The visible units $\boldsymbol{s}\in\{0,1\}^N$ represent measurement outcomes 
in the computational basis, while the hidden units $\boldsymbol{h}\in\{0,1\}^M$ encode correlations between spins.
The interaction weights $\boldsymbol{W} \in \mathbb{R}^{N \times M}$ are shared across all values of $g$,
while the bias parameters are functions of the control parameter.

Conditioned on $g$, the RBM energy is (compare Eq.~\eqref{eq:rbm_energy}):
\begin{equation}
E_\theta(\boldsymbol{s},\boldsymbol{h}\mid g)
= - \sum_i b_i(g)\, s_i
  - \sum_j c_j(g)\, h_j
  - \sum_{i,j} s_i W_{ij} h_j ,
\label{eq:conditional_energy}
\end{equation}

The corresponding free energy $F_\theta(\boldsymbol{s}\mid g)$ and model
distribution $p_\theta(\boldsymbol{s}\mid g)$ are defined analogously to
Eqs.~\eqref{eq:boltzmann_prob_v} and \eqref{eq:rbm_free_energy}. Exploiting
stoquasticity and following Eq.~\eqref{eq:pos_wavefunction_ansatz}, we define the conditional
wavefunction as
\begin{equation}
\Psi_\theta(\boldsymbol{s}\mid g) := \sqrt{p_\theta(\boldsymbol{s}\mid g)} . \label{eq:conditional_wavefunction_ansatz}
\end{equation}

\subsection{Hypernetwork-based conditioning}

\label{sec:hypernetwork}
This construction is conceptually related to conditional \acp{RBM}
\cite{Mnih2011}, where external variables enter the model through linear
modulations of the bias terms. We generalize this idea by using a
hypernetwork to implement a nonlinear modulation of the RBM biases as a
function of the control parameter $g$.
Specifically, a lightweight multilayer perceptron (MLP) takes $g$ as input
and outputs feature-wise scale and shift parameters for the visible and
hidden biases.

Using \ac{FiLM}~\cite{Perez2018}, the biases are
modulated as

\begin{align}
\boldsymbol{b}(g) &= \Big( 1+\bm{\gamma}^{\;b}(g)\Big)\odot \boldsymbol{b}_{\mathrm{base}}
                    + \boldsymbol{\beta}^{b}(g), \\
\boldsymbol{c}(g) &= \Big( 1+\bm{\gamma}^{\;c}(g)\Big) \odot \boldsymbol{c}_{\mathrm{base}}
                    + \boldsymbol{\beta}^{c}(g),
\end{align}
where $\odot$ denotes the Hadamard product. This results in the architecture shown in Fig.~\ref{fig:main}{(b)}.

\subsection{Symmetrized free energy}

In the ferromagnetic phase of the TFIM $(g<J)$, the ground state is doubly degenerate, corresponding to all spins aligning up or down. A standard RBM often breaks this symmetry spontaneously, converging to only one of the two attractors. To strictly enforce the physical $\mathbb{Z}_2$ spin-flip symmetry, we adopt a symmetrized ansatz where the probability is modeled as an equal superposition of a configuration $\boldsymbol{s}$ and its inverse $1-\boldsymbol{s}$:
\begin{align} 
    p_{\text{sym}}(\boldsymbol{s} \mid g) &\propto e^{-E_\theta(\boldsymbol{s}\mid g)} + e^{-E_\theta(1-\boldsymbol{s}\mid g)} \label{eq:symmetrized_prob}\\
    F_{\mathrm{sym}}(\boldsymbol{s}\mid g) &:=-\log\Big(e^{-F_\theta(\boldsymbol{s}\mid g)}+e^{-F_\theta(1-\boldsymbol{s}\mid g)}\Big). \label{eq:symmetrized_free_energy}
\end{align}

This ansatz ensures that $p_{\text{sym}}(\boldsymbol{s})=p_{\text{sym}}(1-\boldsymbol{s})$ by construction. While this symmetrization breaks the conditional independence of the hidden units required for standard block Gibbs sampling, we restore efficient sampling by introducing a latent symmetry variable $u \in \{0,1\}$ into the Markov chain. The full derivation of the symmetrized free energy and the resulting \emph{augmented Gibbs sampling} are detailed in Appendix \ref{app:symmetrized_ansatz}.

\subsection{Training and sampling}
\label{sec:training_sampling}

The model parameters are trained by minimizing the \ac{KL} divergence between the empirical data distribution and $p_\theta(\boldsymbol{s}\mid g)$, following the RBM training procedure reviewed in Sec.~\ref{sec:theory}. Parameter updates are performed using the stochastic gradient learning rule given in Eq.~\eqref{eq:grad_update}, where the model expectation was approximated using \ac{CD-$k$}.

For each minibatch sample $(\boldsymbol{s}, g)$, the positive phase is evaluated at the data configuration $\boldsymbol{s}$, while the negative phase is obtained from $k$ steps of augmented Gibbs sampling scheme described in Appendix~\ref{app:symmetrized_ansatz} conditioned on the same value of $g$. To improve exploration of configuration space, a small fraction of Gibbs chains is initialized from random configurations rather than from data samples.

All RBM and hypernetwork parameters are optimized jointly using the ADAM \cite{kingma_adam_2015} optimizer with an inverse-sigmoid learning-rate schedule.

\section{\MakeUppercase{Evaluation}}
\label{sec:evaluation}

We assess the quality of the learned conditional wavefunction
$\Psi_\theta(\boldsymbol{s}\mid g)$ using two complementary criteria.
First, we evaluate local physical observables that probe operator-level properties of the
state. Second, we consider global state properties such as
fidelity, second-order Rényi entropies, and fidelity susceptibility, which are sensitive to
the full wavefunction structure.
Together, these tests probe both the local physical features and the global
accuracy of the reconstructed state.
All quantities are benchmarked against \ac{ED}.
We focus on system sizes small enough to retain numerically exact reference results, namely $1\times 16$ chain and $L\times L$ lattices with $L\in\{3,4\}$.

\begin{figure*}[htb]
    \centering
    \begin{overpic}[width=\linewidth]{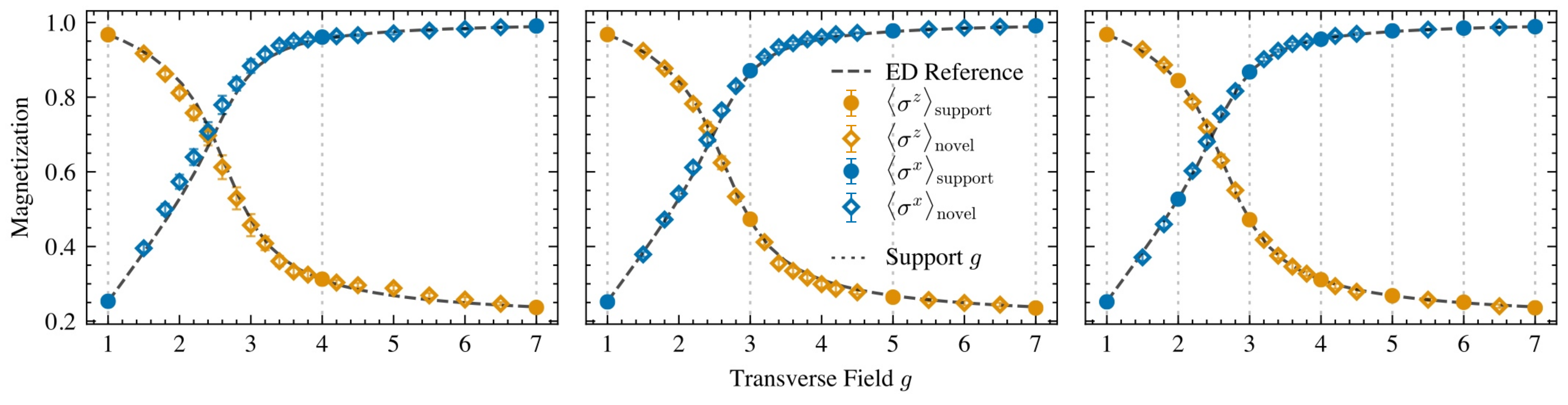}
            \put(3,1.5){\textbf{(a)}}
            \put(35.5,1.5){\textbf{(b)}}
            \put(67.5,1.5){\textbf{(c)}}
    \end{overpic}

    \caption{\textbf{Magnetization reconstruction and interpolation in $g$.}
TFIM on the $4\times4$ lattice: longitudinal magnetization
$\langle\sigma^z(g)\rangle$ (orange) and transverse magnetization $\langle\sigma^x(g)\rangle$ (blue)
as a function of the transverse field $g$.
Gray dashed lines show the exact-diagonalization (ED) reference.
Markers show estimates from the learned conditional RBM state: filled circles are \emph{support} fields used for training,
open diamonds are \emph{novel} fields used only for evaluation.
Vertical dotted lines indicate the support locations.
Panels (a)-(c) correspond to increasingly dense support sets, illustrating that adding support points improves the accuracy of interpolation on novel $g$ values.}

    \label{fig:magnetizations}
\end{figure*}

\subsection{Benchmark I: Local observable estimation from the learned state}
\label{sec:observable_benchmark}

Given a trained model $p_\theta(\boldsymbol{s}\mid g)$, expectation values at
fixed $g$ are estimated by Monte Carlo sampling
$\boldsymbol{s}^{(k)}\sim p_\theta(\cdot\mid g)$.
The wavefunction representation
$\Psi_\theta(\boldsymbol{s}\mid g)\propto \exp[-F_\theta(\boldsymbol{s}\mid g)/2]$
then enables local-estimator expressions based on amplitude ratios.

\subsubsection{Local estimator for general observables}
\label{sec:local_estimator}

For a general operator $\hat O$ with matrix elements
$O_{\boldsymbol{s}'\boldsymbol{s}}=\langle\boldsymbol{s}'|\hat O|\boldsymbol{s}\rangle$,
the expectation value can be written as
\begin{equation}
\langle \hat O\rangle_g
= \sum_{\boldsymbol{s},\boldsymbol{s}'}
\Psi_\theta(\boldsymbol{s}\mid g)\,
O_{\boldsymbol{s}\boldsymbol{s}'}\,
\Psi_\theta(\boldsymbol{s}'\mid g),
\end{equation}
We rewrite this as an expectation over
$p_\theta(\boldsymbol{s}\mid g)=\Psi_\theta^2(\boldsymbol{s}\mid g)$,
\begin{equation}
\langle \hat O\rangle_g
= \sum_{\boldsymbol{s}} p_\theta(\boldsymbol{s}\mid g)\,
O_\theta^{[L]}(\boldsymbol{s}\mid g),
\end{equation}
where the local estimator is
\begin{equation}
O_\theta^{[L]}(\boldsymbol{s}\mid g)
:= \sum_{\boldsymbol{s}'}
O_{\boldsymbol{s}'\boldsymbol{s}}
\frac{\Psi_\theta(\boldsymbol{s}'\mid g)}{\Psi_\theta(\boldsymbol{s}\mid g)} .
\end{equation}
Using $\Psi_\theta\propto e^{-F_\theta/2}$, the amplitude ratio reduces to
\begin{equation}
\frac{\Psi_\theta(\boldsymbol{s}'\mid g)}{\Psi_\theta(\boldsymbol{s}\mid g)}
= \exp\!\left[-\tfrac12\big(F_\theta(\boldsymbol{s}'\mid g)
- F_\theta(\boldsymbol{s}\mid g)\big)\right],
\end{equation}
so the partition function $Z_\theta$ cancels in the ratio and the local estimator is tractable provided that $\hat O$ is sparse in the computational basis, i.e. for each $\boldsymbol{s}$ only a small number of configurations $\boldsymbol{s}'$ yield nonzero $O_{\boldsymbol{s}'\boldsymbol{s}}$. This condition is satisfied by local and few-body observables. For dense operators the evaluation of $O_\theta^{[L]}$ becomes exponentially costly. This reformulation can be viewed as an instance of importance sampling, where configurations $\boldsymbol{s}$ are sampled from $p_\theta(\boldsymbol{s}\mid g)=|\Psi_\theta(\boldsymbol{s}\mid g)|^2$ and the local estimator $O_\theta^{[L]}(\boldsymbol{s}\mid g)$ plays the role of the corresponding importance-weighted integrand.

Diagonal observables are recovered as a special case. Since only
$\boldsymbol{s}'=\boldsymbol{s}$ contributes, the amplitude ratio simplifies to
unity and the local estimator reduces to $O_\theta^{[L]}(\boldsymbol{s}\mid g)
= O(\boldsymbol{s})$. The expectation value then takes the simple form
\begin{equation}
\langle \hat O\rangle_g
= \sum_{\boldsymbol{s}} p_\theta(\boldsymbol{s}\mid g)\, O(\boldsymbol{s}) \approx \frac{1}{n_{\mathrm{MC}}}\sum_{k=1}^{n_{\mathrm{MC}}} O(\boldsymbol{s}^{(k)}).
\end{equation}
which is estimated by Monte Carlo averaging over samples from $\boldsymbol{s}^{(k)}\sim p_\theta(\cdot\mid g)$.

\subsubsection{TFIM observables}
\label{sec:tfim_observables}

For the transverse-field Ising model, we evaluate the longitudinal magnetization
\begin{equation}
\langle \sigma^z(g)\rangle := \frac{1}{N}\sum_{i=1}^N \langle \sigma_i^z\rangle_g, \label{eq:magnetization_z}
\end{equation}
which reflects the change in spin alignment as the transverse field is varied. The same diagonal-estimator framework also applies to two-point correlators $\langle \sigma_i^z \sigma_j^z\rangle_g$, but we do not analyze them further and restrict attention to magnetizations.

Off-diagonal observables are accessed via the amplitude ratios as introduced in Sec.~\ref{sec:local_estimator}.
In particular, the transverse magnetization can be written as
\begin{equation}
\langle \sigma^x(g)\rangle := \frac{1}{N}\sum_{i=1}^N \langle \sigma_i^x\rangle_g.
\end{equation}
Equivalently, this quantity corresponds to measuring $\sigma^z$ after a local Hadamard rotation on each spin, since $H \sigma^z H = \sigma^x$.
In the local-estimator picture, each contribution therefore involves a single spin-flipped configuration relative to $\boldsymbol{s}$. 

We report both $\langle \sigma^z(g)\rangle$ and $\langle \sigma^x(g)\rangle$ to assess whether the reconstructed state
reproduces the expected behavior across the phase diagram, see Fig.~\ref{fig:magnetizations}.

\begin{figure}[htb]
    \centering
    \includegraphics[width=\linewidth]{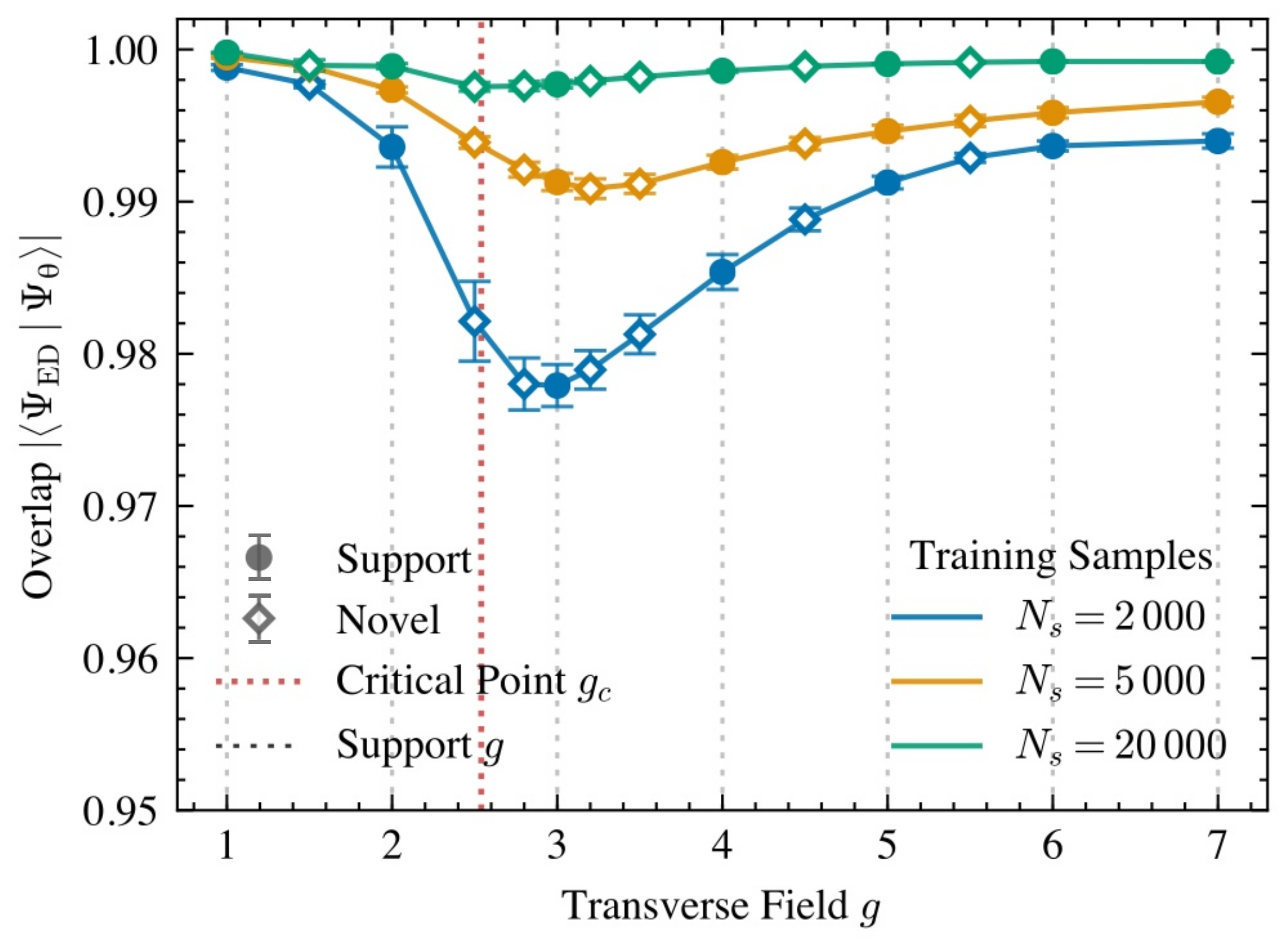}
    \caption{\textbf{Exact overlap and sample efficiency.}
Exact overlap $|\langle \Psi_{\mathrm{ED}}(g)\mid \Psi_\theta(g)\rangle|$
between the ED ground state and the reconstructed RBM state as a function of the transverse field $g$ (here for the $4\times4$ TFIM).
Colors denote the number of training samples per support point, $N_s\in\{2\,000,5\,000,20\,000\}$.
Filled circles indicate \emph{support} field strengths used for training, while open diamonds indicate \emph{novel} fields strengths used only for evaluation.
Gray vertical dotted lines mark the support locations, and the red dotted line marks the finite-size critical field $g_c$.
The overlap remains close to unity throughout the sweep, with the largest deviations concentrated near the crossover region and a systematic improvement as $N_s$ increases.}

    \label{fig:overlap}
\end{figure}

\begin{figure}[htb]
    \centering
    \includegraphics[width=\linewidth]{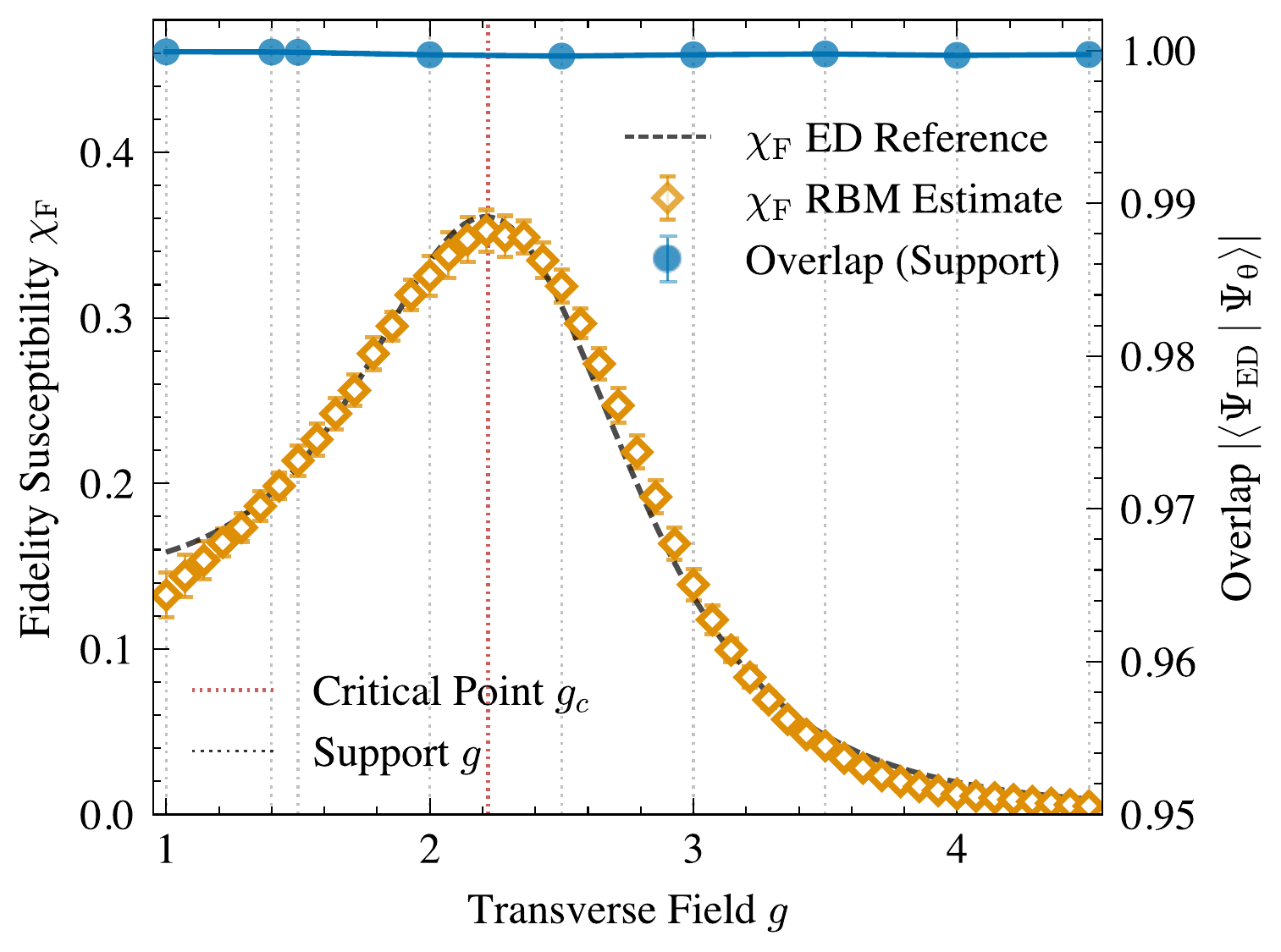}
    \caption{\textbf{Fidelity susceptibility and exact overlap.}
Fidelity susceptibility $\chi_F(g)$ of the $3\times3$ TFIM as a function of the transverse field $g$.
Orange diamonds show the RBM estimate obtained from variance of the free energy gradients (see Sec.~\ref{sec:fidelity_susceptibility}), while the gray dashed curve is the ED reference.
The blue curve (right axis) reports the ED-RBM overlap on the training support points, demonstrating near-unity state fidelity across all field strengths.
Vertical gray dotted lines indicate the support field strengths, and the red dotted line marks the finite-size critical field $g_c$.
The reconstructed $\chi_F(g)$ reproduces the ED peak in the crossover region. }

    \label{fig:susceptibility_smaller}
\end{figure}

\subsection{Benchmark II: Global state-level comparison with exact diagonalization}
\label{sec:ed_benchmark}

For this comparison, we deliberately focus on small lattice sizes so that direct, numerically exact benchmarks against the ground state $|\Psi_{\mathrm{ED}}(g)\rangle$ obtained by \ac{ED} are possible. This provides a gold standard for state-level
comparisons that does not rely on variational or approximate reference states.

\subsubsection{Overlap and fidelity}
\label{sec:fidelity}

We quantify agreement between learned quantum state representation and the underlying true ground state via the fidelity
\begin{equation}
\mathcal{F}_\theta(g)
= \left|\langle \Psi_{\mathrm{ED}}(g)\mid \Psi_\theta(g)\rangle\right|^2 .
\end{equation}
Since both states can be chosen real-valued and non-negative in the computational
basis, the overlap reduces to an explicit sum,
\begin{equation}
\langle \Psi_{\mathrm{ED}}(g)\mid \Psi_\theta(g)\rangle
= \sum_{\boldsymbol{s}}
\Psi_{\mathrm{ED}}(\boldsymbol{s}\mid g)\,
\frac{\exp\!\left[-\tfrac12 F_\theta(\boldsymbol{s}\mid g)\right]}
{\sqrt{Z_\theta(g)}},
\end{equation}
with $Z_\theta(g)=\sum_{\boldsymbol{s}}\exp[-F_\theta(\boldsymbol{s}\mid g)]$.
For the system sizes considered here, $Z_\theta(g)$ is evaluated exactly,
yielding exact fidelities (up to floating point accuracy). 

For larger systems, the overlap can instead be estimated by Monte Carlo
sampling.
Rewriting the sum as an expectation value with respect to
$p_\theta(\boldsymbol{s}\mid g)$ yields
\begin{align}
\langle \Psi_{\mathrm{ED}}(g)\mid \Psi_\theta(g)\rangle
&= \sum_{\boldsymbol{s}} p_\theta(\boldsymbol{s}\mid g)\,
\frac{\Psi_{\mathrm{ED}}(\boldsymbol{s}\mid g)}
{\Psi_\theta(\boldsymbol{s}\mid g)} \nonumber \\
&= \mathbb{E}_{p_\theta(\boldsymbol{s}\mid g)}
\!\left[
\frac{\Psi_{\mathrm{ED}}(\boldsymbol{s}\mid g)}
{\Psi_\theta(\boldsymbol{s}\mid g)}
\right],
\end{align}
where the ratio involves the normalized amplitudes $\Psi_\theta(\boldsymbol{s}\mid g)$.
For large systems, obtaining this normalization requires estimating the partition function $Z_\theta(g)$, which can be approximated using annealed importance sampling \cite{salakhutdinov_quantitative_2008}.

In practice, this estimator can suffer from large variance especially for low overlap NQS, since it involves
an approximate evaluation of $Z_\theta(g)$. This makes \ac{MC} overlap estimation challenging for large systems.
\ac{NQS} representations with tractable or directly normalized amplitudes, such as autoregressive or recurrent architectures \cite{hibat-allah_recurrent_2020}, may alleviate this issue and provide more stable overlap estimators, but exploring such approaches is left for future work.

\subsubsection{Fidelity susceptibility}
\label{sec:fidelity_susceptibility}

To probe the parametric response of the learned family of states, we compute the
fidelity susceptibility, defined as the second derivative of the logarithmic
fidelity with respect to a small parameter shift,
\begin{align}
\chi_F(g)
&:= -\left.\frac{\partial^2}{\partial \delta^2}
\log \mathcal{F}_\theta(g,g+\delta)\right|_{\delta=0}.
\end{align}

Expanding the perturbed state $|\Psi_\theta(g+\delta)\rangle$ to second order in
$\delta$ and evaluating the normalized overlap, one finds that all terms
involving second derivatives cancel. Following standard derivations
\cite{zanardi_information-theoretic_2007,gu_fidelity_2010}, this yields the equivalent
expression
\begin{equation}
\chi_F(g)
=
\frac{\langle \partial_g \Psi_\theta(g)\mid \partial_g \Psi_\theta(g)\rangle}
{\langle \Psi_\theta(g)\mid \Psi_\theta(g)\rangle}
-
\frac{|\langle \Psi_\theta(g)\mid \partial_g \Psi_\theta(g)\rangle|^2}
{\langle \Psi_\theta(g)\mid \Psi_\theta(g)\rangle^2}.
\label{eq:fidelity_susceptibility_standard}
\end{equation}
In practice, rather than resorting to numerical finite differences, we exploit the differentiability of the HyperRBM ansatz. As we derive in Appendix~\ref{app:fidelity_susceptibility_derivation}, Eq.~\eqref{eq:fidelity_susceptibility_standard} can be rewritten as the variance of the gradients of the free energy:
\begin{equation}
\chi_F(g) = \frac{1}{4} \mathrm{Var}_{\boldsymbol{s} \sim p_\theta(\cdot\mid g)} \left[ \frac{\partial F_\theta(\boldsymbol{s}\mid g)}{\partial g} \right].
\end{equation}
This formulation allows us to evaluate $\chi_F(g)$ efficiently using automatic differentiation. We sample a batch of configurations from the model at a fixed $g$, compute the gradients $\partial_g F_\theta$ for each sample via backpropagation and calculate the sample variance. This directly assesses the coherence of the learned parametric family and enables the identification of the quantum critical point without prior knowledge of its location.

\subsubsection{Rényi entropy via the replica trick}\label{sec:renyi_entropy}

To quantify the non-local correlations captured by the reconstructed state $|\Psi_\theta(g)\rangle$, we examine the second Rényi entropy of a spatial subsystem $A$,
\begin{align}
    S_2(A) := -\log \Tr(\rho_A^2), \\ \rho_A := \Tr_{A^\perp} \big(|\Psi_\theta(g)\rangle\langle\Psi_\theta(g)|\big),
\end{align}
which can be estimated via the \emph{replica trick}. We briefly recap the derivation below for completeness following Refs.~\cite{hastings_measuring_2010,torlai2018augmenting}. 

\paragraph{Two-copy observable and the Swap operator.}
The core insight of the replica trick is that non-linear functionals of a state $\rho$ (such as purity, $\Tr(\rho^2)$) can be expressed as linear expectation values over a tensor product of identical copies, $\rho \otimes \rho$. 
Related two-copy constructions also appear in quantum learning and shadow tomography, where joint measurements on $\rho^{\otimes 2}$ enable efficient estimation of Pauli observables
\cite{chen_optimal_2024,king_triply_2025,huang_information-theoretic_2021,huang_quantum_2022,tran_one_2025,noller_infinite_2025}.
While those protocols typically employ Bell-basis measurements, the same two-copy principle underlies the swap-based purity estimator used here.

In our context, we employ the spatial Swap operator, $\mathrm{Swap}_A$, which acts on the computational basis of the two replicas $\boldsymbol{s}^{(1)}$ and $\boldsymbol{s}^{(2)}$ by exchanging their configurations within subsystem $A$:
\begin{align}
    \mathrm{Swap}_A \, | \boldsymbol{s}^{(1)}_A, \boldsymbol{s}^{(1)}_{A^\perp} \rangle \otimes | \boldsymbol{s}^{(2)}_A, \boldsymbol{s}^{(2)}_{A^\perp} \rangle 
    = 
    | \boldsymbol{s}^{(2)}_A, \boldsymbol{s}^{(1)}_{A^\perp} \rangle \otimes | \boldsymbol{s}^{(1)}_A, \boldsymbol{s}^{(2)}_{A^\perp} \rangle.
\end{align}
Using the identity $\Tr(\rho_A^2) = \Tr\left( (\rho \otimes \rho) \, \mathrm{Swap}_A \right)$, the purity becomes an expectation value over the joint state of the two replicas:
\begin{align}
    \Tr(\rho_A^2) = \langle \Psi_\theta(g) | \otimes \langle \Psi_\theta(g) | \; \mathrm{Swap}_A \; | \Psi_\theta(g) \rangle \otimes | \Psi_\theta(g) \rangle.
\end{align}

\paragraph{RBM Estimator and Free Energies.}
A significant advantage of combining the replica trick with Energy-Based Models like RBMs is the cancellation of normalization constants. 
For a fixed conditioning input $g$, the RBM wavefunction $\Psi_\theta(\boldsymbol{s}\mid g)$ is defined as in Eq.~\eqref{eq:conditional_wavefunction_ansatz}, and all quantities below are implicitly conditioned on $g$.
When we expand the purity expectation value in the computational basis, the intractable partition functions in the numerator and denominator cancel perfectly. We are left with an estimator relying solely on the free energies $F_\theta$:
\begin{align}
    \Tr(\rho_A^2) &= \sum_{\boldsymbol{s}^{(1)},\boldsymbol{s}^{(2)}} p_\theta(\boldsymbol{s}^{(1)}) \, p_\theta(\boldsymbol{s}^{(2)}) \, R_A(\boldsymbol{s}^{(1)},\boldsymbol{s}^{(2)}),
\end{align}
where samples are drawn from the joint distribution $p_\theta(\boldsymbol{s}^{(1)})p_\theta(\boldsymbol{s}^{(2)})$ and the \textit{swap ratio} $R_A$ is given by:
\begin{align}
    R_A(\boldsymbol{s}^{(1)},\boldsymbol{s}^{(2)}) 
    &= \frac{ \Psi_\theta(\boldsymbol{s}^{(2)}_A,\boldsymbol{s}^{(1)}_{A^\perp}) \, \Psi_\theta(\boldsymbol{s}^{(1)}_A,\boldsymbol{s}^{(2)}_{A^\perp}) }{ \Psi_\theta(\boldsymbol{s}^{(1)}) \, \Psi_\theta(\boldsymbol{s}^{(2)}) } \nonumber \\
    &= \exp \left[ -\frac{1}{2} \left( 
        \Delta F_{\text{swap}}(\boldsymbol{s}^{(1)}, \boldsymbol{s}^{(2)}) 
    \right) \right],
\end{align}
with the free energy difference
\begin{align}
    \Delta F_{\text{swap}} = F_\theta(\boldsymbol{s}^{(2)}_A,\boldsymbol{s}^{(1)}_{A^\perp}) + F_\theta(\boldsymbol{s}^{(1)}_A,\boldsymbol{s}^{(2)}_{A^\perp}) - F_\theta(\boldsymbol{s}^{(1)}) - F_\theta(\boldsymbol{s}^{(2)}).
\end{align}
Thus $\Tr(\rho_A^2)=\langle R_A\rangle$ with
$\boldsymbol{s}^{(1)},\boldsymbol{s}^{(2)}\sim p_\theta$ independently, which allows us to estimate $S_2(A) = -\log \langle R_A \rangle$ using standard Monte Carlo sampling without ever computing the partition function. In practice, this naive estimator can exhibit large variance for increasing
subsystem size. One can alleviate this by following the improved \emph{ratio trick} as described in
Refs.~\cite{hastings_measuring_2010,torlai2018augmenting} (see also the
Supplementary Material of Ref.~\cite{Torlai2017}).


\begin{figure*}[htb]
    \centering
    \begin{subfigure}[t]{0.495\linewidth}
        \centering
        \begin{overpic}[width=0.87\linewidth]{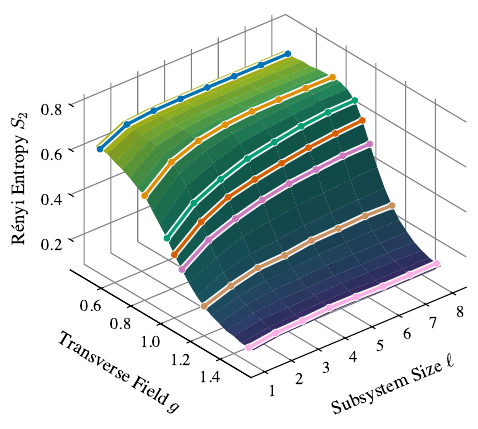}
            \put(3,3){\textbf{(a)}}
        \end{overpic}
        \phantomsubcaption\label{fig:entropy_a}
    \end{subfigure}\hfill
    \begin{subfigure}[t]{0.495\linewidth}
        \centering
        \begin{overpic}[width=\linewidth]{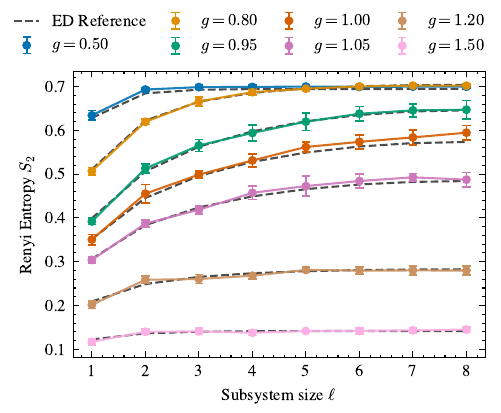}
            \put(3,3){\textbf{(b)}}
        \end{overpic}
        \phantomsubcaption\label{fig:entropy_b}
    \end{subfigure}

    \caption{\textbf{Second R\'enyi entropy across the \ac{TFIM} transition.}
    TFIM on the $1\times16$ chain: second R\'enyi entropy $S_2(\ell,g)$ of a contiguous subsystem of size $\ell$ (Sec.~\ref{sec:renyi_entropy}).
    \textbf{(a)} RBM estimate evaluated on a dense grid ($\Delta g=0.05$) on control parameter $g$, revealing a steep gradient around $g\approx1$ that is captured smoothly as a function of $g$.
    Colored overlay curves indicate the training support fields strengths.
    \textbf{(b)} Overlaid $S_2(\ell)$ slices at different support fields $g$: solid colored curves (markers) are RBM estimates and black dashed curves show the exact-diagonalization (ED) reference.
    The RBM reproduces both the transition-associated drop and the expected entropy scale, saturating near $S_2\approx\ln 2$ for this finite system.}
    \label{fig:entropy}
\end{figure*}

\section{\MakeUppercase{Experimental Setup}}
\label{sec:experimental_setup}

We evaluate the  HyperRBM ansatz on the \ac{TFIM} geometries defined in Sec.~\ref{sec:tfim_setup}, specifically a $1 \times 16$ chain and $L \times L$ square lattices with $L \in \{3,4\}$. We start with baseline magnetization benchmarks, then vary the number of training samples per support point to study sample efficiency, and finally use the best-performing setup for the subsequent experiments. Unless stated otherwise, experiments are repeated over 10 independent random seeds, and we report the median with one standard deviation.

\subsection{Training data and supports}

At each support point $g$, we generated a pool of $20\,000$ projective measurements in the computational basis, sufficient to cover all experiments.

The support points were chosen to span the relevant phase transitions for each geometry:
\begin{itemize}
    \item \textbf{1D Chain:} We chose support field strengths spanning $g\in[0.5,1.5]$ to cover the critical point at $g_c\approx1.0$.
    \item \textbf{2D Lattices:} We explored a broader range $g \in [1.0, 7.0]$, covering the ferromagnetic and paramagnetic phases as well as the critical region.
\end{itemize}
For reference, \ac{ED} studies on finite 2D lattices with periodic boundary conditions report finite-size critical fields $g_c$ \cite{yu_fidelity_2009}, defined as the locations where the fidelity susceptibility peaks. For the $3\times3$ and $4\times4$ cases, we use $g_c \approx 2.22$ and $g_c \approx 2.54$, respectively.

For overlap evaluation, we stored the exact ground states $\ket{\Psi_{\mathrm{ED}}(g)}$ at all support points and at additional $g$ values used to test interpolation.

\subsection{Model architectures and protocols}

We use three closely related training setups, corresponding to the different benchmarks considered below. All hyperparameters are summarized in Appendix~\ref{app:hyperparams}.
Across all benchmarks, we use a HyperRBM with the symmetrized free-energy ansatz (Sec.~\ref{sec:methods}) to enforce the global $\mathbb{Z}_2$ symmetry.

\paragraph{Magnetization interpolation.}
For the magnetization results in Sec.~\ref{sec:bench1_mag}, 
we train models on the $4\times4$ lattice over $g\in[1,7]$. The boundary points $g=1$ and $g=7$ are always included, while the number of intermediate support points is progressively increased.

\paragraph{Sample efficiency.}
We study sample efficiency using $N_s\in\{2\,000,\,5\,000,\,20\,000\}$ samples per support point. The resulting overlap--vs--$g$ curves and their dependence on $N_s$ are reported in Fig.~\ref{fig:overlap} and discussed in Sec.~\ref{sec:bench2_fidelity_sample_eff}.

\paragraph{Fidelity susceptibility.}
Training uses uniformly spaced support points for the $3\times3$ and $4\times4$ lattices, plus one extra support point at the lower end of the range. We estimate $\chi_F(g)$ from reconstructed states using the free-energy-gradient variance estimator described in Sec.~\ref{sec:fidelity_susceptibility}. For each field value $g$, we draw $10^5$ Monte Carlo samples for estimation. Results are shown in Fig.~\ref{fig:susceptibility_smaller} and Fig.~\ref{fig:main}(c). For the $3\times3$ case, the former also shows the \ac{ED}--\ac{RBM} overlap on a secondary axis.

\paragraph{Second R\'enyi entropy.}
For the $1\times16$ chain, we evaluate $S_2(\ell,g)$ on a dense grid with $\Delta g=0.05$. For each field strength, we perform extensive sampling using up to $3\times10^5$ model samples and $20$ Gibbs steps. This allows us to get smooth results despite relying on the naive swap estimator (Sec.~\ref{sec:renyi_entropy}). Results are reported in Fig.~\ref{fig:entropy} (Sec.~\ref{sec:results_renyi}).


\section{\MakeUppercase{Results}}
\label{sec:results}
\subsection{Benchmark I: Magnetizations and interpolation in \texorpdfstring{$g$}{g}}\label{sec:bench1_mag}

\label{sec:results_observable_benchmark}

We first benchmark the learned conditional state using local observables, focusing on the transverse and longitudinal magnetizations $\langle \sigma^x(g)\rangle$ and $\langle \sigma^z(g)\rangle$. 

To assess interpolation in the control parameter $g$, we evaluate the trained model on a fixed grid of $21$ field values spanning $g\in[1,7]$, including both training support points and unseen values. This evaluation is repeated while progressively increasing the number of support points, as shown in Fig.~\ref{fig:magnetizations} from left to right.

Across all support densities, the reconstructed magnetizations reproduce the expected physical behavior: the longitudinal magnetization decreases monotonically with increasing $g$, while $\langle \sigma^x(g)\rangle$ increases and approaches unity at large fields. The most pronounced variation occurs in the finite-size crossover region, where both observables change rapidly in opposite directions. Deep in the paramagnetic regime, $\langle \sigma^x(g)\rangle$ saturates near $1$, while the longitudinal magnetization approaches a small finite-size plateau.

Notably, even with only three support points, the reconstructed magnetizations already show good agreement with the exact curves (Fig.~\ref{fig:magnetizations}(a)). Increasing the support density primarily refines the reconstruction in the finite-size crossover region.

\subsection{Benchmark II: Global state-level properties}
\label{sec:bench2_global_state}

We next assess the learned wavefunction at the state level using global properties, including overlap, fidelity susceptibility, and the second-order R\'enyi entropy.

\subsubsection{Exact overlap and sample efficiency} \label{sec:bench2_fidelity_sample_eff}
Figure~\ref{fig:overlap} shows the overlap
$|\langle \Psi_{\mathrm{ED}}(g)\mid \Psi_\theta(g)\rangle|$ on the $4\times4$
lattice for different numbers of training samples per support point.
The overlap remains close to one across the full range of $g$, with the
largest deviations occurring in the crossover region near the finite-size
critical field $g_c$. Even for $N_s=2\,000$ samples per support point, the
fidelity exceeds $99\%$ outside the crossover region and remains above
$98\%$ at its minimum. Increasing the dataset size systematically reduces this dip and
improves interpolation at novel $g$ values, while the overlap
rapidly saturates outside the crossover region for all training regimes.

Based on this behavior, we use $N_s=20\,000$ samples per support point for the fidelity-susceptibility calculations, where higher precision is desired.

\subsubsection{Fidelity susceptibility from reconstructed states}\label{sec:bench2_fid_suscep}
Figure~\ref{fig:susceptibility_smaller} reports the fidelity susceptibility
$\chi_F(g)$ on the $3\times3$ lattice, computed via the variance of the free-energy gradients.
The \ac{RBM} estimate follows the \ac{ED} reference and reproduces the peak in the
crossover region near $g_c$, suggesting that the conditional model captures the parametric response. At the same time, the \ac{ED}--\ac{RBM} overlap
on the support points remains near unity (blue curve), confirming that the
susceptibility agreement is consistent with high global state fidelity.

Figure~\ref{fig:main}(c) demonstrates similar accuracy on the $4\times4$ lattice, where the model correctly identifies the susceptibility peak around the critical field $g_c\approx2.54$.

\subsubsection{Second R\'enyi entropy: non-local correlations and interpolation in \texorpdfstring{$g$}{g}}
\label{sec:results_renyi}
Figure~\ref{fig:entropy} benchmarks the learned conditional family using the second R\'enyi entropy $S_2(\ell,g)$ on the $1\times16$ chain.
Evaluating the \ac{RBM} on a dense grid of fields yields the surface shown in Fig.~\ref{fig:entropy_a}. 
The smoothness of the surface demonstrates that the model successfully interpolates the wavefunction continuously in $g$.
Consequently, the conditional state faithfully resolves the steep gradient in entanglement entropy around the critical point $g\approx1$ without requiring point-wise retraining.

Figure~\ref{fig:entropy_b} details the corresponding $S_2(\ell)$ slices at different support fields $g$ overlaid in a single panel.
The \ac{RBM} estimates closely follow the \ac{ED} reference across all cuts and reproduce the expected entropy scale, saturating near $S_2\approx\ln 2$ in the ferromagnetic phase.
We observe the largest deviations at the largest subsystems, consistent with the increased variance of swap-based estimators for large $\ell$ (Sec.~\ref{sec:renyi_entropy}).

\section{\MakeUppercase{Conclusion \& Outlook}}
\label{sec:conclusion_outlook}
We introduced a \acf{pQST} framework based on a hypernetwork-conditioned \ac{RBM} (HyperRBM) for ground states of the one- and two-dimensional transverse-field Ising model. The transverse field strength $g$ acts as a continuous control parameter and modulates the \ac{RBM} biases via FiLM. On a $1\times 16$ chain and $L\times L$ for $L \in \{3,4\}$ lattices, we showed that a single conditional model for each lattice size can represent a continuous family of ground states across the phase diagram. The learned states reproduce key physical signatures, including transverse and longitudinal magnetizations, already with modest training data. With sufficient samples, the HyperRBM achieves high overlap with the exact ground states (exceeding $99\%$) across both phases and remains accurate in the critical region. In addition, we accurately estimate second-order Rényi entropies for the 1D chain using extensive sampling, indicating that the approach captures non-local correlations and entanglement structure. Finally, treating the reconstructed states as a coherent parametric family, we extract the fidelity susceptibility for 2D lattices and identify the quantum critical point without prior knowledge of its location or an explicit order parameter.
All results were obtained with modest computational resources, relying exclusively on CPU-based training and sampling (details in Appendix~\ref{app:compute}).

In this work, system sizes were limited by the use of exact diagonalization for ground-truth fidelities and susceptibility benchmarks. A natural next step is to combine HyperRBMs with scalable data generation techniques, such as quantum Monte Carlo or tensor-network methods, to extend parametric tomography to larger lattices. 

Beyond the TFIM, an important next step is to extend parametric tomography to non-stoquastic Hamiltonians. This can be achieved by combining the present conditioning scheme with the general complex-valued \ac{RBM} ansatz introduced in Ref.~\cite{Torlai2017}, where separate networks parameterize amplitudes and phases. In this setting, the hypernetwork would condition both amplitude and phase
parameters, enabling parametric learning of genuinely complex-valued quantum states. Preliminary experiments indicate that the approach extends beyond stoquastic models when combined with complex-valued \ac{RBM} ansatzes, although a systematic study is left for future work.

Besides improving sample and training efficiency, this structured conditioning also supports interpretability analyses. Because physical control parameters act through a low-dimensional modulation pathway, one can inspect how changes in the Hamiltonian translate into structured updates of model parameters and how these responses evolve across phases or near critical points. More broadly, it remains an open question whether hypernetwork-equipped \ac{NQS} can provide a flexible framework for learning physically meaningful manifolds of quantum states, including thermal or parameter-dependent dynamical states.

\section*{\MakeUppercase{Code and Data Availability}}
The code used to generate the results and the datasets analyzed in this study will be made available at 
\ifcamera
\url{https://github.com/simontonner/hyperrbm-parametric-qst}. 
\cite{tonner_hyper_2025}
\else
\url{https://github.com/real_repo_in_camera_ready_version}.
\fi

\ifcamera
\section*{\MakeUppercase{Acknowledgements}}

AI tools such as ChatGPT (GPT-5 series) and Gemini (Gemini 3 Pro) were used to refine phrasing in parts of the manuscript and to assist with figure styling. All results and final wording were reviewed and validated by the authors.

The team at JKU Linz acknowledges funding from the European Research Council (ERC) via the Starting grant q-shadows (101117138) and from the Austrian Science Fund (FWF) via the SFB BeyondC (10.55776/FG7).
\fi

\begingroup
\small
\printbibliography
\endgroup

\appendix
\section{\MakeUppercase{Appendix}}

\subsection{Symmetrized ansatz and augmented sampling}
\label{app:symmetrized_ansatz}

\subsubsection{Symmetrized Free Energy}
In the ferromagnetic phase of the TFIM ($g < J$), the ground state is doubly degenerate. To strictly enforce the physical $\mathbb{Z}_2$ spin-flip symmetry, we define the model probability as a mixture of the standard \ac{RBM} energy $E_\theta(\boldsymbol{s})$ and the energy of the flipped configuration $E_\theta(1-\boldsymbol{s})$. The effective free energy $F_{\text{sym}}(\boldsymbol{s})$ is:
\begin{equation}
F_{\text{sym}}(\boldsymbol{s}) = -\ln \left( e^{-F_\theta(\boldsymbol{s})} + e^{-F_\theta(1-\boldsymbol{s})} \right),
\end{equation}
where $F_\theta(\boldsymbol{s})$ is the standard \ac{RBM} free energy. The term for the flipped configuration is computed efficiently using the identity $(1-\boldsymbol{s})^{\mathsf{T}} \boldsymbol{W} = \boldsymbol{\Sigma}_W^{\mathsf{T}} - \boldsymbol{s}^{\mathsf{T}} \boldsymbol{W}$, where $\boldsymbol{\Sigma}_W$ is the vector of column sums of $\boldsymbol{W}$.

\subsubsection{Augmented Gibbs Sampling}
Standard block Gibbs sampling is inapplicable because the symmetrized mixture distribution breaks the conditional independence of hidden units. We employ a data augmentation scheme \cite{tanner_calculation_1987} by introducing a binary latent variable $u \in \{0,1\}$ ($u=0$: normal, $u=1$: flipped). Conditioned on $u$, independence is restored.

The sampling procedure is a three-step alternating Gibbs chain. We define $\mathrm{sigmoid}(x) = (1+e^{-x})^{-1}$.

\paragraph{1. Sample Hidden Units ($\boldsymbol{s}, u \to \boldsymbol{h}$).}
Calculate the effective input $\tilde{\boldsymbol{s}} = \boldsymbol{s}$ if $u=0$, or $\tilde{\boldsymbol{s}} = 1-\boldsymbol{s}$ if $u=1$. Sample hidden units independently:
\begin{equation}
p(h_j=1 \mid \boldsymbol{s}, u) = \mathrm{sigmoid}\left( c_j(g) + \sum_i \tilde{s}_i W_{ij} \right).
\end{equation}

\paragraph{2. Sample Symmetry Variable ($\boldsymbol{s}, \boldsymbol{h} \to u'$).}
We compare the energy of the current configuration against its flipped counterpart. Define the energy difference $\Delta E = E_\theta(1-\boldsymbol{s}, \boldsymbol{h}) - E_\theta(\boldsymbol{s}, \boldsymbol{h})$. The probability of switching to the flipped sector is:
\begin{equation}
p(u'=1 \mid \boldsymbol{s}, \boldsymbol{h}) = \mathrm{sigmoid}(-\Delta E).
\end{equation}

\paragraph{3. Sample Visible Units ($\boldsymbol{h}, u' \to \boldsymbol{s}'$).}
Generate a prototype vector $\boldsymbol{s}_{\text{proto}}$ using the standard \ac{RBM} conditional $p(s_i=1 \mid \boldsymbol{h}) = \mathrm{sigmoid}(b_i(g) + \sum_j W_{ij} h_j)$. Apply the symmetry transform:
\begin{equation}
\boldsymbol{s}' = 
\begin{cases} 
\boldsymbol{s}_{\text{proto}} & \text{if } u'=0 \\
1-\boldsymbol{s}_{\text{proto}} & \text{if } u'=1.
\end{cases}
\end{equation}

\subsection{Compute resources}
\label{app:compute}

All experiments were conducted on CPU on an Apple MacBook Pro (M1 Pro) with 16\,GB RAM. Except for runs with extensive Monte Carlo sampling, neither memory nor CPU utilization approached saturation. No individual training or sampling run exceeded 30 minutes.

\subsection{Derivation of fidelity susceptibility via variance}
\label{app:fidelity_susceptibility_derivation}
Here we derive the connection between the fidelity susceptibility and the variance of the free energy gradients used in Sec.~\ref{sec:fidelity_susceptibility}.

We start from the geometric definition of the fidelity susceptibility given in Eq.~\eqref{eq:fidelity_susceptibility_standard}:
\begin{equation}
\chi_F(g)
=
\frac{\langle \partial_g \Psi_\theta\mid \partial_g \Psi_\theta\rangle}
{\langle \Psi_\theta\mid \Psi_\theta\rangle}
-
\left( \frac{\langle \Psi_\theta\mid \partial_g \Psi_\theta\rangle}
{\langle \Psi_\theta\mid \Psi_\theta\rangle} \right)^2,
\tag{\ref{eq:fidelity_susceptibility_standard}}
\end{equation}
where we have suppressed the dependency on $g$ for brevity and assumed real-valued wavefunctions.
We express the wavefunction gradients using the log-derivative identity $\partial_g \Psi_\theta(\boldsymbol{s}) = \Psi_\theta(\boldsymbol{s}) \partial_g \ln \Psi_\theta(\boldsymbol{s})$.

The first term corresponds to the second moment. By explicitly writing out the inner product sum and grouping the normalization factor $\langle \Psi_\theta | \Psi_\theta \rangle$, we recover the model probability $p_\theta(\boldsymbol{s}) = |\Psi_\theta(\boldsymbol{s})|^2 / \langle \Psi_\theta | \Psi_\theta \rangle$:
\begin{align}
\frac{\langle \partial_g \Psi_\theta \mid \partial_g \Psi_\theta \rangle}{\langle \Psi_\theta \mid \Psi_\theta \rangle} 
&= \frac{1}{\langle \Psi_\theta \mid \Psi_\theta \rangle} \sum_{\boldsymbol{s}} \left( \Psi_\theta(\boldsymbol{s}) \frac{\partial \ln \Psi_\theta(\boldsymbol{s})}{\partial g} \right)^2 \nonumber \\
&= \sum_{\boldsymbol{s}} \underbrace{\frac{|\Psi_\theta(\boldsymbol{s})|^2}{\langle \Psi_\theta \mid \Psi_\theta \rangle}}_{p_\theta(\boldsymbol{s})} \left( \frac{\partial \ln \Psi_\theta(\boldsymbol{s})}{\partial g} \right)^2 \nonumber \\
&= \mathbb{E}_{p_\theta} \left[ \left( \partial_g \ln \Psi_\theta \right)^2 \right].
\end{align}

Similarly, the second term corresponds to the squared first moment. The denominator cancels against the state amplitudes in the inner product to again yield the probability weight $p_\theta(\boldsymbol{s})$:
\begin{align}
\frac{\langle \Psi_\theta \mid \partial_g \Psi_\theta \rangle}{\langle \Psi_\theta \mid \Psi_\theta \rangle} 
&= \frac{1}{\langle \Psi_\theta \mid \Psi_\theta \rangle} \sum_{\boldsymbol{s}} \Psi_\theta(\boldsymbol{s}) \left( \Psi_\theta(\boldsymbol{s}) \frac{\partial \ln \Psi_\theta(\boldsymbol{s})}{\partial g} \right) \nonumber \\
&= \sum_{\boldsymbol{s}} \underbrace{\frac{|\Psi_\theta(\boldsymbol{s})|^2}{\langle \Psi_\theta \mid \Psi_\theta \rangle}}_{p_\theta(\boldsymbol{s})} \frac{\partial \ln \Psi_\theta(\boldsymbol{s})}{\partial g} \nonumber \\
&= \mathbb{E}_{p_\theta} \left[ \partial_g \ln \Psi_\theta \right].
\end{align}

The fidelity susceptibility is the difference between these two moments, which is exactly the variance:
\begin{align}
\chi_F(g) &= \mathbb{E}_{p_\theta}\big[ (\dots)^2 \big] - \big( \mathbb{E}_{p_\theta}[\dots] \big)^2 \\
&= \mathrm{Var}_{\boldsymbol{s}\sim p_\theta} \left[ \frac{\partial \ln \Psi_\theta(\boldsymbol{s})}{\partial g} \right].
\end{align}

For the \ac{RBM} ansatz $\Psi_\theta(\boldsymbol{s}) = e^{-F_\theta(\boldsymbol{s})/2} / \sqrt{Z_\theta}$, the log-derivative is:
\begin{equation}
\frac{\partial \ln \Psi_\theta}{\partial g} = -\frac{1}{2}\frac{\partial F_\theta(\boldsymbol{s})}{\partial g} - \frac{1}{2}\frac{\partial \ln Z_\theta}{\partial g}.
\end{equation}
Since the partition function term $\partial_g \ln Z_\theta$ is independent of the configuration $\boldsymbol{s}$, it acts as an additive constant. Because the variance is translation invariant ($\mathrm{Var}[X+c] = \mathrm{Var}[X]$), this term vanishes, yielding the final expression:
\begin{equation}
\chi_F(g) = \frac{1}{4} \mathrm{Var}_{\boldsymbol{s}\sim p_\theta} \left[ \partial_g F_\theta(\boldsymbol{s}) \right].
\end{equation}

\subsection{Hyperparameter configuration}
\label{app:hyperparams}
\begin{strip}
\centering
\captionsetup{type=table}
\caption{\textbf{Table A.1} | Hyperparameter settings across experimental tasks.}
\label{tab:hyperparams}

\vspace{3mm}

\small
\setlength{\tabcolsep}{5pt}
\renewcommand{\arraystretch}{1.15}

\begin{tabularx}{\textwidth}{lXXXX}
\toprule
\textbf{Parameter} & \textbf{Magnetization} & \textbf{Sample Efficiency} & \textbf{Fidelity \newline Susceptibility} & \textbf{R\'enyi Entropy} \\
\midrule

\multicolumn{5}{l}{\textbf{System}} \\
Geometry              & 2D ($4 \times 4$)               & 2D ($4 \times 4$)              & 2D ($4 \times 4$ \& $3 \times 3$)              & Chain ($L=16$) \\
Train. Supports        & var.\ $\left(g \in [1.0, 7.0]\right)$       & $7$ $\left(g \in [1.0, 7.0]\right)$    & $9$ $\left(g \in [1.0, 4.5]\right)$       & $7$ $\left(g \in [0.5, 1.5]\right)$ \\
Train. Samples / Supp. & $20k$                      & $2k$ / $5k$ / $20k$           & $20k$                     & $20k$ \\
\addlinespace[0.4em]

\multicolumn{5}{l}{\textbf{Architecture}} \\
Hidden Units          & $64$                             & $64$                           & $64$                           & $64$ \\
Hypernet. Width     & $64$                             & $64$                           & $64$                           & $64$ \\
\addlinespace[0.4em]

\multicolumn{5}{l}{\textbf{Training}} \\
Epochs                & $50$                             & $50$                           & $50$                           & $50$ \\
Batch Size            & $1024$                           & $1024$                         & $1024$                         & $1024$ \\
CD-$k$ Steps          & $10$                            & $10$                           & $20$                           & $20$ \\
Chain Init.            & Data + $10\%$ Noise             & Data + $10\%$ Noise            & Data + $10\%$ Noise            & Data + $10\%$ Noise \\
Weight Init. (STD)     & $0.01$                           & $0.05$                          & $0.01$                         & $0.01$ \\
LR Sigmoid Schedule   & $10^{-2} \to 10^{-4}$            & $10^{-2} \to 10^{-4}$          & $10^{-2} \to 10^{-4}$          & $10^{-2} \to 10^{-4}$ \\
\addlinespace[0.4em]

\multicolumn{5}{l}{\textbf{Evaluation}} \\
Method                & MC Sampling                      & Exact via $Z_\theta$           & MC Sampling                    & MC Sampling \\
Eval. Samples          & $1k$                            & N/A                            & $5k$                          & $20k$ \\
Eval. CD-$k$ Steps          & $10$                            & N/A                            & $20$                           & $20$ \\
\bottomrule
\end{tabularx}
\end{strip}
\par\mbox{}\vspace{1em} 

\end{document}